# Tunable inverse topological heterostructure utilizing $(Bi_{1-x}In_x)_2Se_3$ and multi-channel weak-antilocalization effect

Matthew J. Brahlek[1,†,#], Nikesh Koirala[1,†], Jianpeng Liu[1], Tahir I. Yusufaly[1], Maryam Salehi[2], Myung-Geun Han[3], Yimei Zhu[3], David Vanderbilt[1], and Seongshik Oh[1,*]

[1]Department of Physics & Astronomy, Rutgers, The State University of New Jersey, Piscataway, New Jersey 08854, U.S.A.

[2]Department of Materials Science and Engineering, Rutgers, The State University of New Jersey, Piscataway, New Jersey 08854, U.S.A.

[3]Condensed Matter Physics & Materials Science, Brookhaven National Lab, Upton, NY 11973, U.S.A.

[†]These authors contributed equally to this work

[#]Present address: Department of Materials Science and Engineering, Pennsylvania State University, University Park, Pennsylvania 16802, U.S.A.

*ohsean@physics.rutgers.edu

**Abstract:** **In typical topological insulator (TI) systems the TI is bordered by a non-TI insulator, and the surrounding conventional insulators, including vacuum, are not generally treated as part of the TI system. Here, we implement the first material system where the roles are reversed, and the TSS form around the non-TI (instead of the TI) layers. This is realized by growing a layer of the tunable non-TI $(Bi_{1-x}In_x)_2Se_3$ in between two layers of the TI $Bi_2Se_3$ using the atomically-precise molecular beam epitaxy technique. On this tunable inverse topological platform, we systematically vary the thickness and the composition of the $(Bi_{1-x}In_x)_2Se_3$ layer and show that this tunes the coupling between the TI layers from strongly-coupled metallic to weakly-coupled, and finally to a fully-decoupled insulating regime. This system can be used to probe the fundamental nature of coupling in TI materials and provides a tunable insulating layer for TI devices.**

The topological classification scheme is rooted in the ability to distil the global properties of an object into a single number known as a topological invariant. This notion of topology can be extended from the archetypal example of geometric topology, where shapes are classified based solely on the number of holes, to electronic materials, where the main physical implication occurs on the boundary between materials that belong to different topological classes. In the 3-dimensional (3D) class of topological insulators (TIs) unusual topological surface states (TSS) form on the 2-dimensional (2D) boundary with non-TIs. These surface states have metallic, gapless energy bands, which disperse linearly with momentum like photons, and the spins of the surface electrons are locked perpendicular to the direction of their momentum (see refs. [1,2]). TSS have been experimentally confirmed by various surface sensitive probes such as angle-resolved photo emission spectroscopy [3–5], scanning tunneling microscopy [6], and more recently, transport measurements [7–16].

The TSS are an entirely interfacial phenomenon, which form across the interface between a TI and a trivial insulator. As shown in Fig. 1(a), experiments to probe the nature of the TSS thus far have viewed them as forming at the interface between a single-slab TI and a trivial insulator such as vacuum. However, as depicted in Fig. 1(b), equivalent TSS will form at the interfaces when the role of the trivial insulator and the TI are reversed. Such an inverted structure is physically attainable, and provides new opportunities that cannot be accessed in existing TI materials. Using the atomic precision of the molecular beam epitaxy (MBE) technique, we report the first implementation of such a system by inserting a layer of a tunable non-TI $(Bi_{1-x}In_x)_2Se_3$ (bandgap ~1.3 eV at $x = 100\%$ [17–20]) between two layers of the TI $Bi_2Se_3$ [21,22], as shown in Fig. 1(c). This tunable inverse topological (TIT) system allows us to investigate how the conducting channels interact through the bulk $(Bi_{1-x}In_x)_2Se_3$ layer in a regime far beyond what is accessible in any existing TI materials: from a strong insulator with a bandgap of ~1.3 eV, which is far greater than those (~0.3 eV) of existing TIs, all the way down to a zero-gap semimetal. It is also important to note that $(Bi_{1-x}In_x)_2Se_3$ is unique as a component for TI heterostructures in that it is the only non-TI material that shares the same crystal structure with a TI: this is critical for the formation of atomically smooth interfaces as seen in Fig. 1(c).

By varying the thickness of the non-TI barrier layer from the ultra-thin (1 quintuple layer, 1 QL ≈ 1 nm) to the thick regime (> 100 QL), we find that the $Bi_2Se_3$ layers become electrically isolated at an $In_2Se_3$ thickness of ~3 QL, as shown by transport measurements, and first-principles calculations indicate that this coincides with the emergence of the TSS. Further, by decreasing $x$ in the $(Bi_{1-x}In_x)_2Se_3$ barrier layer from $x$ = 100% to 30%, the conduction band offset (potential barrier height) is lowered, and the coupling strength gradually increases for a fixed barrier layer thickness. For a fixed composition $x$, the $Bi_2Se_3$ layers undergo a coupled-to-decoupled transition as the thickness grows beyond a critical value. However, when $x$ is reduced below ~30%, the barrier layer becomes metallic and the system remains fully coupled over the entire thickness range.

The weak anti-localization (WAL) effect is a common feature of the magneto-resistance in strongly spin-orbit-coupled 2D systems such as TI thin films (see Supplemental Materials for measurement geometry, and note that both layers of $Bi_2Se_3$ are electrically contacted using In contacts) [10,12,13,15,16,23–28]. As shown in Fig. 2(a), the WAL effect is typically seen as a sharp cusp at small field in resistance vs magnetic field, which is fit by the Hikami-Larkin-Nagaoka (HLN) formula $\Delta G(B) = -\tilde{A}e^2/(2\pi h)(ln[\hbar/(4el_\phi^2 B)] - \Psi[1/2 + \hbar/(4el_\phi^2 B)])$, where $h$ is Planck's constant, $e$ is the electron charge, $\Psi$ is the digamma function, and the two fitting parameters are $\tilde{A}$, a constant, and $l_\phi$, the de-phasing length [29]. In general $l_\phi$ is limited by the inelastic scattering length, which depends strongly on microscopic details such as disorder and phonons (see Fig. S6 in Supplemental Materials). In contrast, $\tilde{A}$ has been found to be much more robust. In single slab $Bi_2Se_3$ layers it has been found that $\tilde{A} \approx 1$ due to the conducting bulk state, which mediates electrical coupling between the surface transport channels in the film. However, recent studies show that $\tilde{A}$ can increase to 2 if the top and bottom surface channels in TI films are electrically isolated from each other [12,13,15,16,23]. These observations show that $\tilde{A}$ represents the number of decoupled 2D conducting channels in strongly spin-orbit-coupled systems, thus providing a means to probe the coupling strength between adjacent 2D channels. Therefore, as we show below, the TIT heterostructure allows us to take two adjacent $Bi_2Se_3$ layers, each with $\tilde{A} \approx 1$, and determine their electric coupling by tracking $\tilde{A}$. Using this, we fully map out the dependence of the interlayer coupling on both composition and thickness of the $(Bi_{1-x}In_x)_2Se_3$ barrier layer.

By introducing an $In_2Se_3$ layer in the center of a 60-QL $Bi_2Se_3$ slab (*i.e.* $Bi_2Se_3$-$In_2Se_3$-$Bi_2Se_3$ with thicknesses 30-*t*-30 QL) induces an $\tilde{A} = 1 \rightarrow 2$ transition with increasing $In_2Se_3$ thickness *t*, as shown in Fig. 2(b), which implies that beyond a critical thickness of the $In_2Se_3$ layer, the top and bottom TI layers become decoupled. Due to the large bandgap of $In_2Se_3$, ~1.3 eV, compared with ~0.3 eV for $Bi_2Se_3$, the top and bottom $Bi_2Se_3$ layers become electrically isolated for barrier thicknesses as small as ~3 QL, whereas a similar transition occurs only above ~20 QL of separation between the two surfaces in bulk insulating single-slab $Bi_2Se_3$ films [15,16,26] and no such transition occurs in the commonly available bulk-metallic single-slab $Bi_2Se_3$ films [26]. Figure 2(c) shows an extension of this experiment where another unit of $Bi_2Se_3$-$In_2Se_3$ has been added ($Bi_2Se_3$-$In_2Se_3$-$Bi_2Se_3$-$In_2Se_3$-$Bi_2Se_3$ with corresponding thicknesses of 30-20-30-*t*-30 QL); $\tilde{A}$ responds by transitioning from $2 \rightarrow 3$ with increasing $In_2Se_3$ thickness, which confirms and extends the counting nature of the $\tilde{A}$ parameter.

To understand the microscopic origin of this transition, we carried out first-principles calculations based on density-functional theory (DFT). We first performed calculations on bulk $Bi_2Se_3$ and $In_2Se_3$, which were extended to $Bi_2Se_3$-$In_2Se_3$ supercells by allowing the construction of Wannierized effective Hamiltonians (see Supplemental Materials for more details). Figure 2(d) shows that the bandgap at the $\Gamma$ point near the interface of $Bi_2Se_3$-$In_2Se_3$ closes with increasing $In_2Se_3$ thickness as the wavefunction overlap between the interfacial states dies out. The spatial electronic properties can be further seen by tracking the real space density of the states around the Fermi level (RDOS, see ref. [30] and Supplemental Materials), as shown in Fig. 2(e-h). This calculation shows that the RDOS increases near the $Bi_2Se_3$-$In_2Se_3$ interface even for a single QL of $In_2Se_3$, indicating a new state has begun to emerge. By 2-3 QL of $In_2Se_3$, the RDOS splits, peaking near the $Bi_2Se_3$-$In_2Se_3$ interfaces and diminishing near the center, which implies the formation of the gapless interfacial states and the development of an insulating bulk state in the middle of $In_2Se_3$. The finite density of states in the $In_2Se_3$ region is due to the evanescent decay of the TSS wavefunctions into the $In_2Se_3$ layer, and extends around ~2 QL into the $In_2Se_3$ layer, which is better seen for the relatively thick 8 QL $In_2Se_3$ in Fig. 2(h). This shows that the emergence of the interface states from the DFT calculation coincides with the WAL $\tilde{A} = 1 \rightarrow 2$ transition, both of which suggest that the two $Bi_2Se_3$ layers are fully isolated beyond ~3 QL of $In_2Se_3$.

We can extend the ability to engineer and explore how the $\tilde{A} = 1 \rightarrow 2$ transition evolves by mixing Bi into the $In_2Se_3$ barrier layer, which controls the insulating properties of the barrier layer. It was previously shown that $(Bi_{1-x}In_x)_2Se_3$ undergoes a composition-dependent topological and metal-insulator phase transitions: it first undergoes a TI to non-TI transition near $x \approx 3 - 7\%$ [18–20], then transitions into a weakly insulating variable-range-hopping state near $x \approx 15\%$ and finally into a strong band insulator for $x > 25\%$ [18]. Therefore, by adjusting $x$, we can control the coupling strength between the TI layers. This process is sketched in Fig. 3 (a), and demonstrated by plots of $\tilde{A}$ versus $(Bi_{1-x}In_x)_2Se_3$ thickness in Fig. 3 (b-d), which show that as $x$ decreases from 40 to 30 and to 20%, the transition region is pushed to larger thickness, and Fig. 3(e-g), which show that in $\tilde{A}$ versus composition at fixed thickness the transition occurs at smaller In composition with increasing thickness.

As shown in Fig. 2 (b-c) and Fig. 3 (b-d), the barrier thickness dependence of $\tilde{A}$ is well fit by a smoothed step function, $\tilde{A}(t) = 2 - 1/(1 + e^{2(t-t_0)/l_0})$, which is characterized by the critical transition thickness $t_0$ and the transition width $l_0$. We have plotted the values of the fitting parameters $t_0$ and $l_0$ in Fig. 3(h), and it can be seen that for $x \gtrsim 30\%$, both $t_0$ and $l_0$ are approximately exponential functions of $x$. However, the empirical exponential dependence breaks down below $x = 30\%$. For $x \gtrsim 30\%$, we fit the experimental behavior to $t_0(x) = \tau e^{-x/x_0}$ and $l_0(x) = \lambda e^{-x/x_0}$, resulting in $\tau \approx 90$ nm, $\lambda \approx 60$ nm, and both $t_0$ and $l_0$ consistently yield $x_0 \approx 25\%$, which coincides with the composition where the exponential trend breaks down. In order to see this breakdown more clearly, we generated the red dotted curve for $x = 20\%$ in Fig. 3(d) by extrapolating the exponential behavior to $x = 20\%$ (see the red stars in Fig. 3(h)); this curve clearly deviates from the experimental data for $x = 20\%$. The origin of this transition is likely due to the intrinsically high Fermi level in these materials: in an ideal TI, the Fermi level ($E_{F,Ideal}$) is naturally at the Dirac point, whereas in real materials, the Fermi level is close to but above the bottom of the bulk conduction band ($E_{F,Real}$). Therefore, due to the natural position of the Fermi level, the barrier layer will become metallic when the conduction bands of the Bi$_2$Se$_3$ layers and the (Bi$_{1-x}$In$_x$)$_2$Se$_3$ barrier layer cross, which, as detailed in the Supplemental Materials, occurs near $x \approx 25\%$. This indicates that below $x \approx 25\%$, the insulating behavior of (Bi$_{1-x}$In$_x$)$_2$Se$_3$ fully breaks down, and gives way to a metallic regime, which coincides with the known composition where the band-insulating state dies out in homogeneous (Bi$_{1-x}$In$_x$)$_2$Se$_3$ films [18].

Figure 3(e-g) shows how $\tilde{A}$ changes with $x$ for fixed barrier thicknesses (10, 20 and 30 QL). For each thickness, $\tilde{A}$ transitions from 1 to 2 with increasing $x$. If the empirical exponential dependence of $t_0$ and $l_0$ on $x$ holds, then there should be enough information to generate a curve that fits these data points. Using $\tilde{A}(t) = 2 - \frac{1}{1+e^{2(t-t_0)/l_0}} \to \tilde{A}(t,x) = 2 - \frac{1}{1+e^{2(t-t_0(x))/l_0(x)}}$, where $l_0(x) \approx 60e^{-x/25}$ and $t_0(x) \approx 90e^{-x/25}$ (the numerical values were obtained above) allows the generation of the curves for $\tilde{A}$ vs $x$ (for greater than ~25%) with *no free parameters*. The solid red curves in Fig. 3(e-g) agree well with the experimental data for $x > 25\%$, and this further confirms that the empirical exponential relations for $l_0$ and $t_0$ hold for all $x$ greater than ~25%. Figure 3(i) summarizes the behavior of $\tilde{A}$ as a function of thickness $t$ and composition $x$. The well-defined behavior of $\tilde{A}$ with both $x$ and $t$ suggests the presence of a fundamental underlying mechanism. However, the exponential dependence of the parameters on composition is not yet understood, and further studies will be needed to fully resolve this.

Much like the TSS that form around the bulk state in a TI, we have shown that in our TIT heterostructure system the coupling between the two TIs is determined solely by the properties of the middle (Bi$_{1-x}$In$_x$)$_2$Se$_3$ layer, which transitions from a full insulator to a semi-metal depending on its composition and thickness. Going forward, the unique properties of (Bi$_{1-x}$In$_x$)$_2$Se$_3$ can be utilized as a tunnel barrier and gate dielectric, which are essential components of many TI devices such as spin injection devices, topological tunnel junctions and field effect transistors. First, it is the only material that has a seamless interface with the TI Bi$_2$Se$_3$. Second, its barrier properties are finely adjustable through the Bi/In ratio. As such, we anticipate that this study will stimulate utilization of (Bi$_{1-x}$In$_x$)$_2$Se$_3$ in various TI nanostructures and devices leading to further discoveries and applications. Finally, it is worth noting that the concept of inserting a tunable barrier layer between 2D channels as a way to manipulate their electronic properties can be applied to non-TI systems as well. Comparing and testing how the electronic properties evolve as various 2D channels merge or split while maintaining the 2D nature will be an interesting subject of future studies.

**Acknowledgements**

This work is supported by NSF (DMR-1308142, DMR-1005838, and DMREF-1233349), and Gordon and Betty Moore Foundation's EPiQS Initiative (GBMF4418). The work at Brookhaven National Lab is supported by U.S. Department of Energy, Office of Basic Energy Science, Division of Materials Science and Engineering under Contract number DE-AC02-98CH10886. We acknowledge J. Garlow for TEM sample preparation using focused-ion beam at Center for Functional Nanomaterials, Brookhaven National Lab.

**Figure Captions**

**Figure 1.** (Color online) Schematic and TEM image comparing the single-slab TI vs inverse TI heterostructure. (a-b) Topological surface states form when there is a change in topological invariant ($v$) across an interface (*i.e.* $\Delta v = 1$). (a) Single-slab TI: a topological material (characterized by a topological invariant $v_{TI} = 1$) surrounded by non-topological insulator ($v_{nTI} = 0$), which forms a metallic topological surface state at the interface (indicated by the arrows). (b) Inverse TI geometry formed by inverting the role of non-TI and the topological material, and a nominally identical surface state forms at this interface. (c) High-resolution high-angle annular dark-field scanning transmission electron microscopy image showing the physical realization of the inverse TI geometry (b) in a $Bi_2Se_3$-$In_2Se_3$-$Bi_2Se_3$ heterostructure.

**Figure 2.** (Color online) Weak anti-localization and first principles calculations for $Bi_2Se_3$–$In_2Se_3$–$Bi_2Se_3$ films. (a) The measured change in conductance and corresponding fit to the HLN equation for $Bi_2Se_3$-$In_2Se_3$-$Bi_2Se_3$ with $In_2Se_3$ thickness of 2 and 50 QL (see Supplemental Materials for more data) (b,c) $\tilde{A}$ extracted from the HLN formula plotted versus thickness of the $In_2Se_3$ layer in $Bi_2Se_3$-$In_2Se_3$-$Bi_2Se_3$ (30-*t*-30 QL): the curves are an empirical function defined in the text. (b), and $Bi_2Se_3$-$In_2Se_3$-$Bi_2Se_3$-$In_2Se_3$-$Bi_2Se_3$ (30-20-30-*t*-30 QL) (c). (d) Calculations of the energy gap at the interface of $Bi_2Se_3$-$In_2Se_3$ at the $\Gamma$ point showing that the gap at the Dirac point closes as the $In_2Se_3$ thickness increases. (e-h) Schematic of the heterostructure formed for increasing $In_2Se_3$ thickness alongside the calculated real space density of the states (RDOS) around the Fermi level as a function of thickness, showing the emergence and decoupling of the interfacial states with thickness.

**Figure 3.** (Color online) Composition and thickness-dependent coupling parameters extracted from weak-antilocalization effect for $Bi_2Se_3$-$(Bi_{1-x}In_x)_2Se_3$–$Bi_2Se_3$ films. (a) Schematic showing the $Bi_2Se_3$-$(Bi_{1-x}In_x)_2Se_3$-$Bi_2Se_3$ structure where the coupling can be modulated by changing the barrier thickness or height. (b-d) $\tilde{A}$ for $Bi_2Se_3$-$(Bi_{1-x}In_x)_2Se_3$-$Bi_2Se_3$ with fixed composition while varying thickness, and (e-g) with fixed thickness while varying composition. Symbols are experimental data, and the lines are fits to the empirical

function (see text). (h) Compositional dependence of the fitting parameters extracted from the curves in (b-d) and Fig. 2(b). Here $t_0$ gives the critical thickness for the $\tilde{A} = 1 \to 2$ transition, while $l_0$ is the width of the transition region. (i) Contour plot of the empirical function for $\tilde{A}$ as a function of $x$ and $t$ for $x \gtrsim 25\%$; below $x \approx 25\%$, we took $\tilde{A}$ to be 1 due to the metallic nature of the barrier in this regime.

**Figure 1 (two-column)**

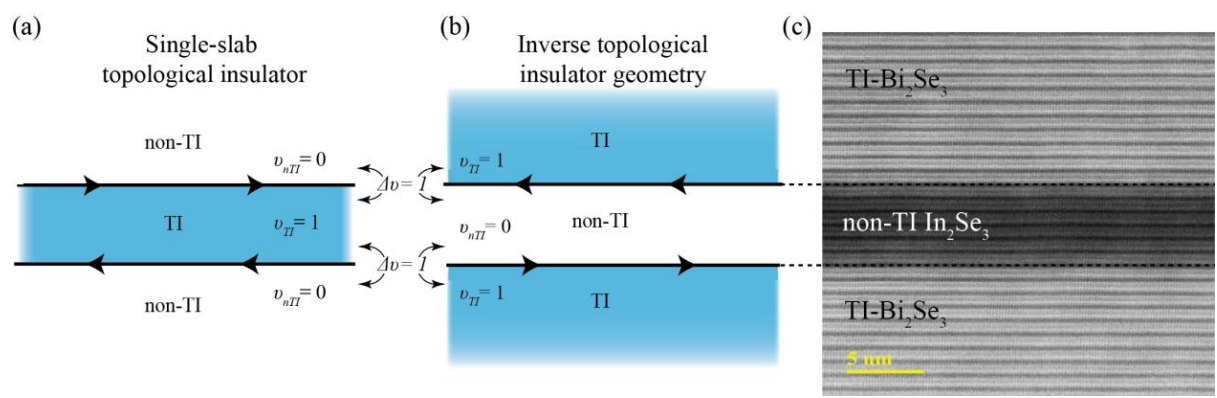

**Figure 2 (two-column)**

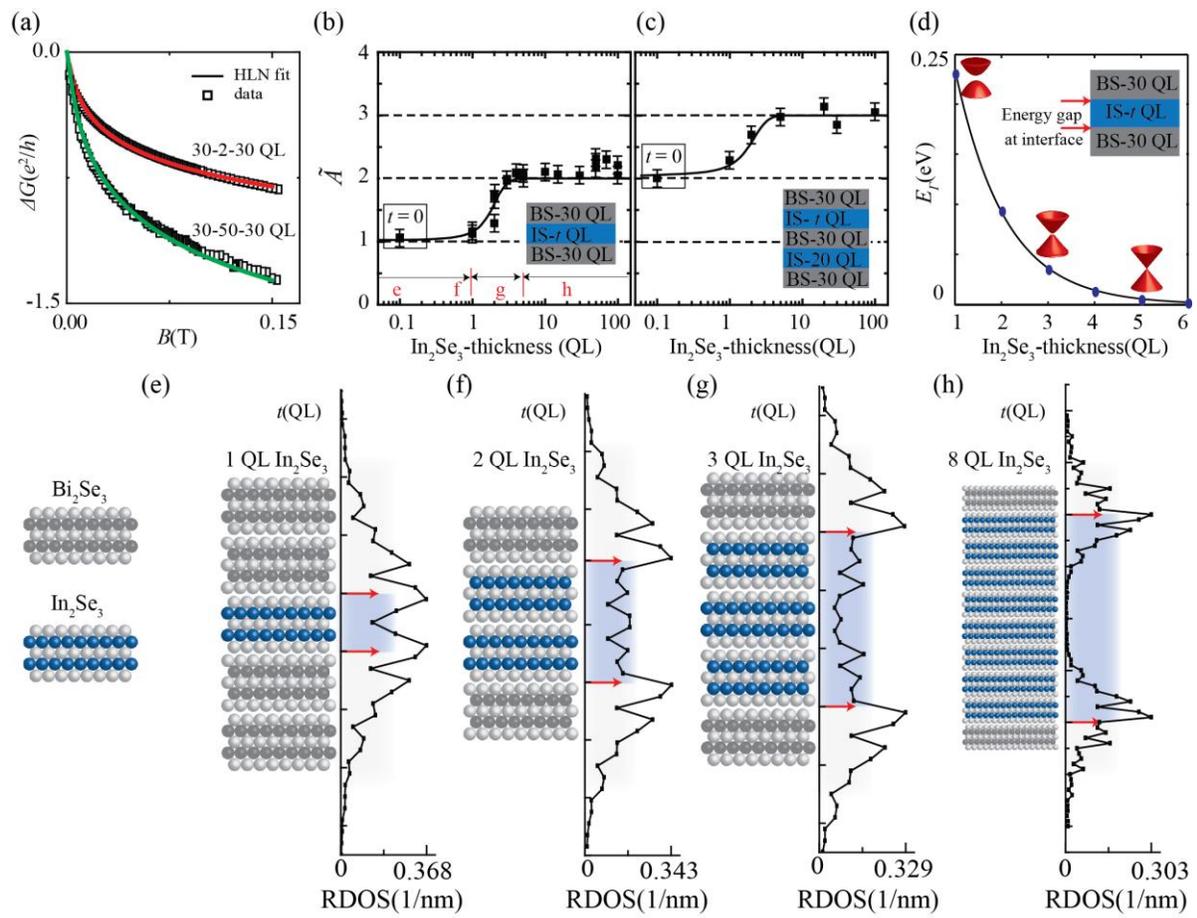

**Figure 3 (two-column)**

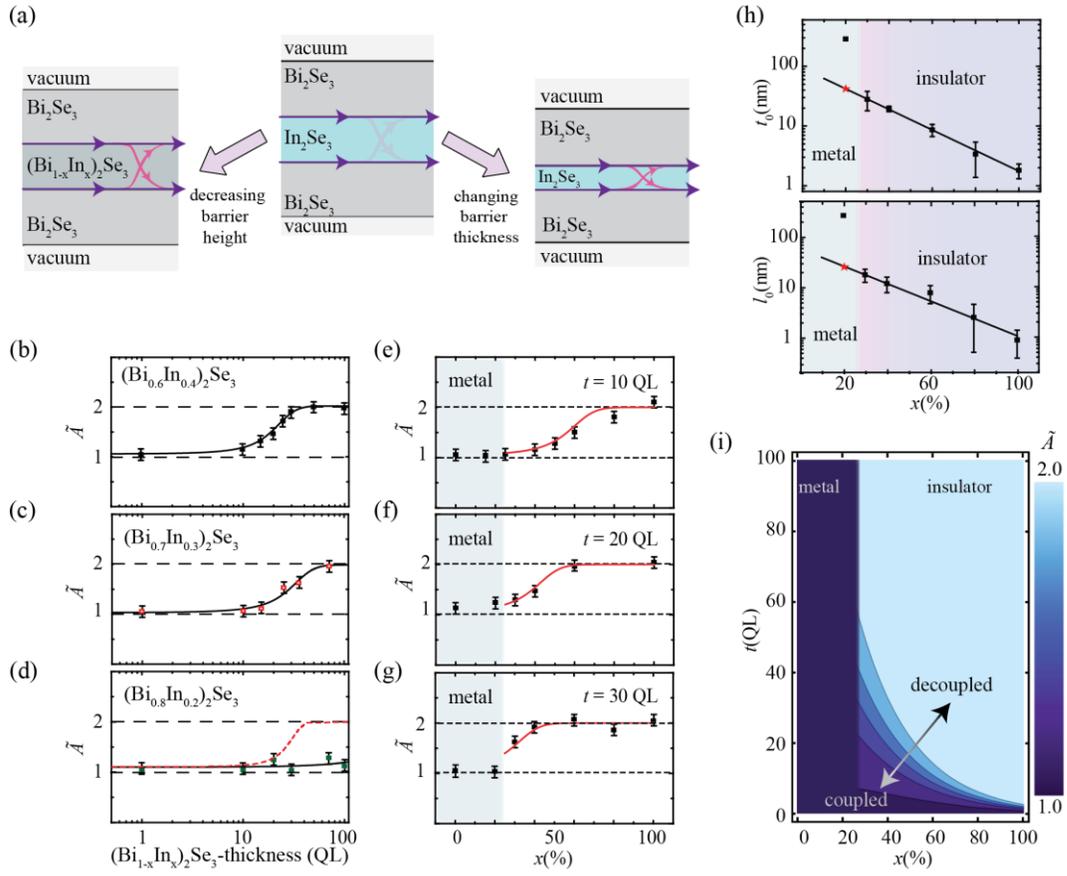

Supplemental Materials for

# Tunable inverse topological heterostructure utilizing $(Bi_{1-x}In_x)_2Se_3$ and multi-channel weak-antilocalization effect

**Contents:**

- **A:** Experimental methods
- **B:** Computational methods
- **C**: Weak anti-localization: numerical fitting

## A: Experimental methods

All samples were grown using 10 mm × 10 mm $c$-plane $Al_2O_3$ substrates. The first $Bi_2Se_3$ layer was grown according to the two-step growth method developed at Rutgers University where the first 3 QL was grown at 135°C, which was followed by slowly annealing the sample to 300°C, where the subsequent 27 QL of $Bi_2Se_3$ layers were grown. Once the first $Bi_2Se_3$ layer finished growth, the $In_2Se_3$ of the specified thickness was grown, followed by the remaining $Bi_2Se_3$ layer. $Bi_2Se_3$ (lattice constant of 4.14 Å) and $In_2Se_3$ (4.00 Å) both fully relax within the first QL of heteroepitaxy. All the samples were then capped by 50 QL of $In_2Se_3$ which stabilized the films during exposure to atmosphere. For the samples with $(Bi_{1-x}In_x)_2Se_3$ as the barrier layer, the same basic recipe was used. The Bi and In cell temperatures were adjusted such that when opened together the resulting film gave the concentration that was sought. All the concentrations were checked by a combination of ex situ Rutherford back scattering spectroscopy and in situ quartz crystal microbalance measurements, and the results were within ±1% of the target values.

TEM sample preparation was carried out with focused-ion beam (FIB) technique using 5 keV $Ga^+$

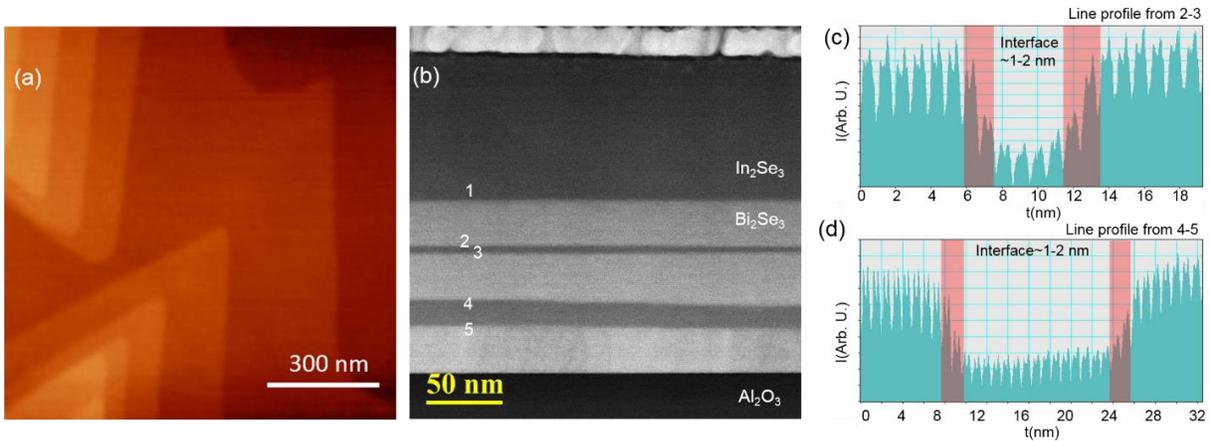

**Figure S1.** (a) 1 μm × 1 μm atomic force microscopy image showing the surface morphology. (b), Wide angle high-resolution high-angle annular dark-field scanning transmission electron microscopy for a 30-20-30-5-30 QL film. (c-d) Electron energy loss spectroscopy profiles taken across the interfaces in (b) as indicated.

ions. A JEOL ARM 200CF equipped with a cold field-emission gun and double-spherical aberration correctors operated at 200 kV was used for high-angle annular dark-field (HAADF) scanning transmission electron microscopy (STEM) with the collection angles ranging from 68 to 280 mrad. As shown by atomic force microscopy and HAADF STEM in Fig. S1 (a-b), the interfaces of the films are flat with well-defined interfaces. Some Bi-In interdiffusion may occur but electron energy loss spectroscopy in Fig. S1 (c-d) shows it to be mostly confined to the first QL of the interface [1].

As shown in Fig. S2, transport measurements were carried out at 1.5 K using the standard 4-point Van der Pauw lead geometry, and the magnetic field was applied perpendicular to the films' surface. Both $Bi_2Se_3$ layers were equally contacted by physically pressing ~mm sized indium contacts into the film. The raw data was symmetrized to remove any odd component from $R_{xx}$ and any even component from $R_{xy}$. The carrier density and mobility of the films ranged between 3-7 × $10^{13}$ /$cm^2$ and 500-1000 $cm^2$/Vs, and there was no correlation between the transport data and the value of $\tilde{A}$. From the WAL fitting, $l_\phi$ ranged between 50-100 nm and also showed little correlation with the other transport data or $\tilde{A}$. The temperature dependence of resistivity for all samples showed typical monotonic decreasing behavior with decreasing temperature, which is typical of a metal. $\tilde{A}$ was independent of temperature below ~20 K, above which deviation occurred as thermal effect suppresses the WAL signal.

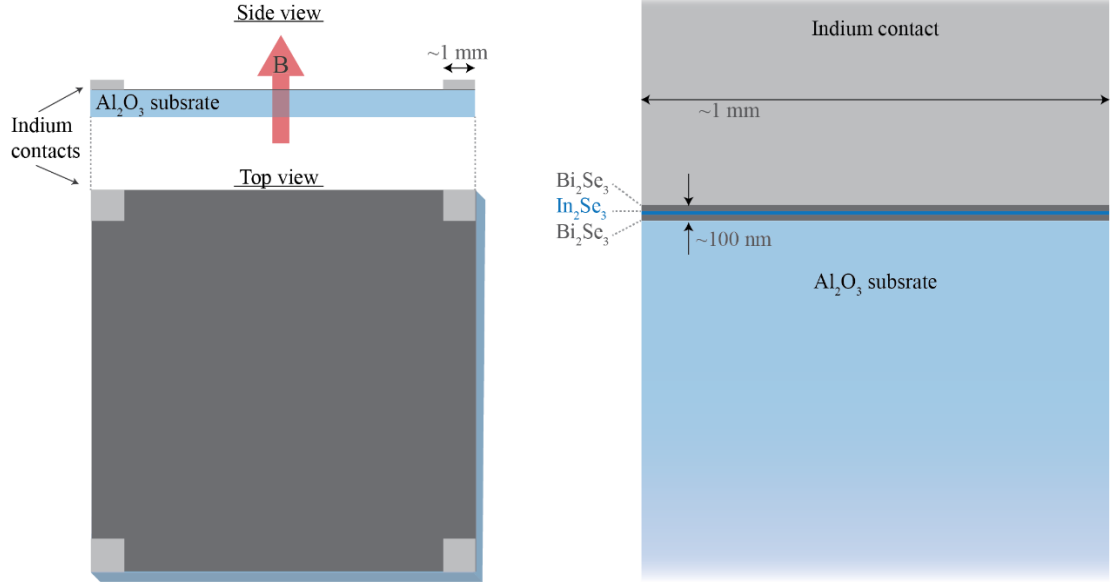

**Figure S2.** Schematic of the experimental setup where the films were grown on 10 mm × 10 mm square substrates, and electrical contact was made using millimeter size indium pads in the Van der Pauw geometry. The blow up on the right shows the scale of the thin film relative to the substrate and contact size (i.e. ~100 nm film relative to millimeter substrate thickness and contact area). The huge aspect ratio (over tens of thousands) between the lateral contact dimensions and the film thickness allows uniform current flow through both the top and bottom TI layers.

# B: Computational methods
## B1. Tunneling between topological interface states

The tunneling between the topological surface states (TSS) in $Bi_2Se_3$-$In_2Se_3$-$Bi_2Se_3$ heterostructures was studied based on density-functional theory (DFT) [2,3]. Calculations on bulk $Bi_2Se_3$ and $In_2Se_3$ were first performed using the Quantum ESPRESSO package [4], with the generalized gradient approximation (GGA) [5] to the exchange-correlation functional and fully relativistic norm-conserving pseudopotentials. The Brillouin zone (BZ) was sampled on an 8×8×8 Monkhorst-Pack [6] $k$ mesh, with an energy cutoff of 55 Ry (1 Ry ≈ 13.6 eV) for $Bi_2Se_3$ and 65 Ry for $In_2Se_3$. The first-principles output was fed into the Wannier90 package to produce Wannier functions (WFs) and to generate a realistic tight-binding (TB) model defined in the chosen Wannier basis [7,8]. 30 Wannier functions were constructed for $Bi_2Se_3$, including all the valence $p$ orbitals, while four extra In 5$s$ orbitals were included for $In_2Se_3$. Both models were constructed in such a way that they exactly reproduce the first-principles bandstructures within a certain energy range, spanning from 3 eV below to 3 eV above the Fermi level.

The supercells including a $Bi_2Se_3$-$In_2Se_3$ interface can be constructed based on the bulk TB models. First, the Wannier-based model Hamiltonians for bulk $Bi_2Se_3$ and $In_2Se_3$, denoted as $H_1$ and $H_2$, were extrapolated to $N_1$ QL and $N_2$ QL slabs stacked in the [111] direction with open boundary conditions. These two isolated slabs were connected together in such a way that all the first-neighbor hopping (here referring to hopping terms between nearest-neighbor QLs) across the interface were taken as the average value of the corresponding hopping terms in the $Bi_2Se_3$ and $In_2Se_3$ bulk TB models. Then the periodic boundary condition was applied to the ($N_1 + N_2$)-QL slab to make it a periodic supercell. The total thickness of Bi$_2$Se$_3$ and In$_2$Se$_3$ was fixed to be $N_1 + N_2 = 12$ QLs, and the thickness of In$_2$Se$_3$ was varied from $N_2 = 1$ to 6 QLs (for the data shown in Fig. 2(h) of the main text, $N_1 + N_2 = 16$ QLs with $N_2 = 8$). Working in the Wannier basis allows for the thickness of In$_2$Se$_3$ in the heterostructure to be highly tunable, and the computational cost is negligible compared with a fully self-consistent interface calculation.

In implementing this procedure, two issues need to be addressed. First, at the bulk level, standard DFT tends to underestimate the energy of the In 5$s$ level. Because the lowest conduction band and highest valence band of In$_2$Se$_3$ are dominated by In 5$s$ and Se 4$p$ orbitals respectively, DFT predicts a smaller band gap compared with experiment [9]. Here a corrective treatment was adopted as described in ref. [9], which involves applying a +0.79 eV rigid shift (taken from many-body GW calculations) to the four In 5$s$ levels in the 34-band model for In$_2$Se$_3$, leaving all the other matrix elements unchanged.

Another issue is the band offset between the two bulk materials. Initially the zeroes of energy of the Wannierized tight-binding models for Bi$_2$Se$_3$ and In$_2$Se$_3$ are inherited from the respective bulk DFT bulk calculations, but as is well known, these are largely arbitrary, as they depend on irrelevant details such as the choice of pseudopotentials. To address this issue, the alignment method based on surface work functions [10] was adopted by carrying out self-consistent surface slab calculations on Bi$_2$Se$_3$ and In$_2$Se$_3$ slabs individually, from which the difference between the average electrostatic potential energy deep in the bulk and in the vacuum was evaluated for each material. This was done by computing the macroscopic-averaged electrostatic potential $\bar{V}(z)$ from the microscopic potential $V(x,y,z)$ as: $\bar{V}(z) = (cA)^{-1} \int_{z-c/2}^{z+c/2} dz \iint_A dxdy\, V(x,y,z)$, where $c$ and $A$ are the cell height (size of a QL) and basal area respectively. For these calculations, a 3-QL slab was used, and slabs were chosen to be separated from each other by a vacuum space of 2.9 nm. The macroscopic averages of the electrostatic potentials are plotted in Fig. S3. Note that due to the non-polar crystal structure and the homogeneous nature of the vacuum $\bar{V}(z)$ remains constant both deep in the bulk and in vacuum. Aligning the vacuum levels, it was concluded that the relative shift between the average electrostatic potential in bulk Bi$_2$Se$_3$ vs In$_2$Se$_3$ is $\Delta V = V_2 - V_1 = 1.776$ eV. Therefore, the arbitrariness in the energy zeroes can be removed by shifting all the Kohn-Sham eigenenergies of In$_2$Se$_3$ using $\tilde{E}_n(\mathbf{k}) = E_n(\mathbf{k}) + \Delta V$.

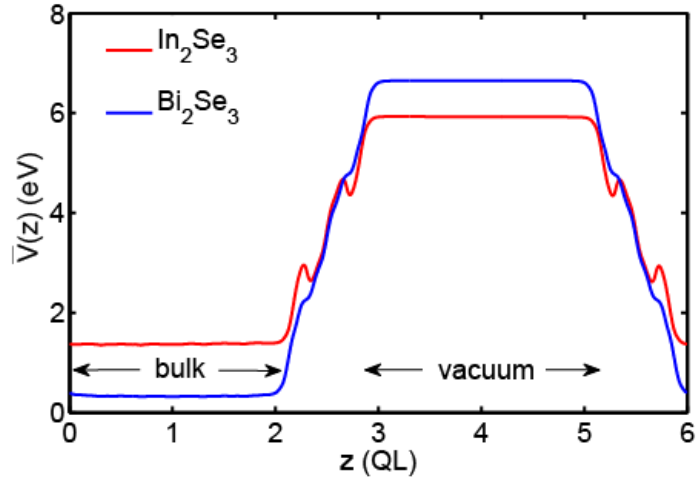

**Figure S3**. The macroscopic average of the electrostatic potentials of Bi$_2$Se$_3$ and In$_2$Se$_3$ slabs.

With the GW correction to In 5$s$ levels and the shift $\Delta V$ on all the In$_2$Se$_3$ on-site energies, the interface model has been well constructed. The eigenvalues were then calculated in the ($k_x$, $k_y$) plane, setting $k_z = 0$. If the TSS do not interact, a doubly degenerate gapless Dirac cone around $\Gamma$ ($k_x = 0$, $k_y = 0$) is expected, but the energy spectrum should become gapped when a tunneling interaction is allowed. Therefore, the band gap at $\Gamma$, denoted as $\Delta(\Gamma)$, should provide a measure reflecting the tunneling amplitude between the TSSs. As shown in Fig. 2(d) in the main text, $\Delta(\Gamma)$ was found to drop exponentially as the thickness of the In$_2$Se$_3$ layer increases. Setting 0.05 eV as a threshold below which the tunneling between the TSS is considered as negligible, the corresponding critical thickness $t_c$ is about ~2.6 QLs, which agrees well with experimental data.

One may also be interested in the real-space distribution of the interface states, which can be easily calculated using the interface model described above. The following quantity is introduced as a weight of the real space density of the interface states around the Fermi level [11] z: $\xi(z) = \sqrt{|\psi_\Gamma^v(z)|^2 + |\psi_\Gamma^c(z)|^2}$, where $\psi_\Gamma^v(z)$ and $\psi_\Gamma^c(z)$ are the components of the Bloch states at $\Gamma$ projected onto the Wannier functions centered at $z$, and the superscripts $v$ and $c$ refer to the highest occupied and lowest unoccupied states respectively. If the Fermi level lies slightly above the conduction band minimum (CBM) at $\Gamma$, $\xi(z)^2$ measures the $z$-dependence of the charge density averaged over the $x$-$y$ plane around the Fermi level. $\xi(z)$ is denoted as the real space density of the states (RDOS) in the main text, as shown in Fig. 2(e-h).

**B2. Band alignment**

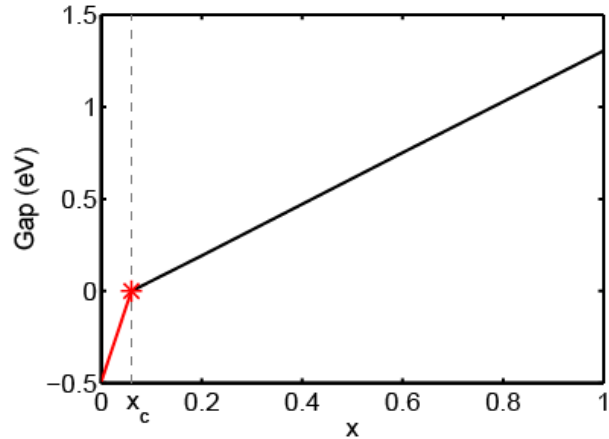

**Figure S4.** The bulk gap of (Bi$_{1-x}$In$_x$)$_2$Se$_3$ at $\Gamma$ from linear interpolations. The asterisk marks the critical point. A negative gap (red segment) indicates a topological band inversion.

The position of the In$_2$Se$_3$ conduction band minimum (CBM) and valence band maximum (VBM) with respect to the Bi$_2$Se$_3$ VBM can also be determined from the above self-consistent slab calculations. It turns out that the In$_2$Se$_3$ CBM and VBM at $\Gamma$ (including the +0.79 eV correction on In 5$s$ levels) are 1.286 eV above and -0.018 eV below the Bi$_2$Se$_3$ VBM respectively. Such information is useful in evaluating the band alignment in (Bi$_{1-x}$In$_x$)$_2$Se$_3$. However, the CBM and VBM positions for different $x$ values cannot be evaluated simply by linearly interpolating the two end points ($x = 0\%$ and $100\%$), because a linear gap-closure picture does not apply to (Bi$_{1-x}$In$_x$)$_2$Se$_3$ over the entire $x$ interval, the bulk band gap vanishes at very low In composition as a result of the In clustering tendency and the presence of In 5$s$ orbitals [9,12,13]. In order to treat the band alignment in (Bi$_{1-x}$In$_x$)$_2$Se$_3$ better, the position of the 3D Dirac point at criticality was also extracted from ref. [9], which is 0.106 eV above the VBM of Bi$_2$Se$_3$. Even though the theoretical critical point of (Bi$_{1-x}$In$_x$)$_2$Se$_3$ ($x_c \approx 16.7\%$) is higher than the experimental value ($x_c \approx 6\%$ [13], $x_c \approx$ 4 − 7% [12]), here it is assumed that the theoretical shift of the 3D Dirac point with respect to the $Bi_2Se_3$ VBM at criticality also applies to the experimental situation. Namely, it is assumed that the 3D Dirac point is 0.106 eV above the $Bi_2Se_3$ VBM at $x = 6\%$.

**Table S2.** Band alignment of $(Bi_{1-x}In_x)_2Se_3$

| $x$ | 0 | 6% | 20% | 60% | 100% |
|---|---|---|---|---|---|
| VBM (eV) | 0 | 0.106 | 0.088 | 0.035 | −0.018 |
| CBM (eV) | 0.490 | 0.106 | 0.280 | 0.786 | 1.286 |

Using the positions of the CBM and VBM at 3 different $x$ values as specified above ($x = 0\%$, 6% and 100%), the CBM and VBM for any other $x$ can be obtained from two separate linear interpolations in the left and right intervals partitioned by $x_c$. Under such an approximation, the gap vs $x$ consists of two linear curves with different slopes, as shown in Fig. S4, instead of a single straight line as predicted by a simple linear-gap-closure picture.

Table S2 and Fig. S5 show the alignments of the CBM and VBM of $(Bi_{1-x}In_x)_2Se_3$ with respect to the VBM of $Bi_2Se_3$ at different $x$. When $x$ is 20%, the CBM of $(Bi_{1-x}In_x)_2Se_3$ is below that of $Bi_2Se_3$, which means that in a realistic case in which the Fermi level is slightly above the CBM of $Bi_2Se_3$, the $(Bi_{1-x}In_x)_2Se_3$ barrier layer would behave as a metal with the TSS extending through the entire barrier layer. On the other hand, the CBM goes above the $Bi_2Se_3$ CBM when $x$ is 60%, such that the $(Bi_{1-x}In_x)_2Se_3$ layer acts as an actual potential barrier which would decouple the two TSS.

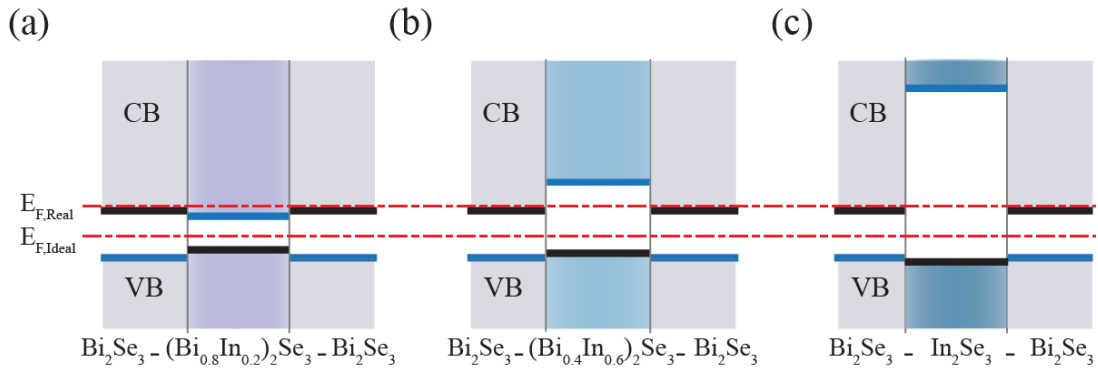

**Figure S5.** Energy bands alignment of $Bi_2Se_3$-$(Bi_{1-x}In_x)_2Se_3$-$Bi_2Se_3$ for $x = 20$, 60 and 100% respectively. For $x \gtrsim 25\%$ the conduction band minimum of $(Bi_{1-x}In_x)_2Se_3$ is above the experimental (real) Fermi level $E_{F,Real}$, which makes the barrier layer insulating, while for $x \lesssim 25\%$ the conduction band minimum drops below $E_{F,Real}$, which makes the barrier metallic: with ideal Fermi levels ($E_{F,Ideal}$), the barrier should remain insulating even for $x = 20\%$.

## C: Weak anti-localization: numerical fitting

As described in the main text, fitting the change in magneto-conductance to the HLN equation, shown in Fig. 2(a) and Fig. S6 (a-f), requires two fitting parameters, the number of conductive channels, $\tilde{A}$, which is the main focus of the main text, and the dephasing length, $l_\phi$, which is plotted in Fig. S6 (g-h) versus thickness. It can be seen that unlike $\tilde{A}$ which shows a very clear dependence on the thickness of the $(Bi_{1-x}In_x)_2Se_3$ barrier layer, $l_\phi$ shows no discernible dependence on the thickness of the barrier layer.

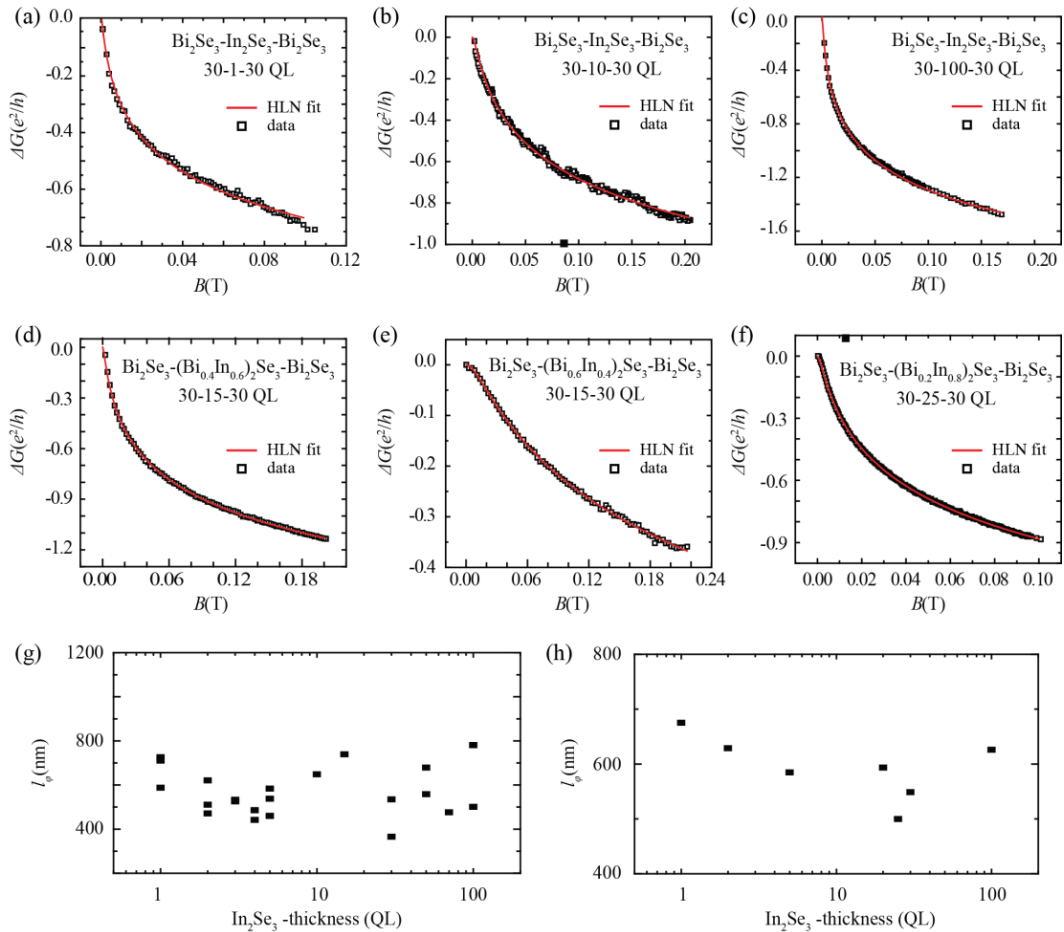

**Figure S6.** Change in conductance and numerical fits to the HLN formula for various barrier thicknesses and compositions (a-f). $l_\phi$ versus $In_2Se_3$ thickness for $Bi_2Se_3$-$In_2Se_3$-$Bi_2Se_3$ (g), and $Bi_2Se_3$-$In_2Se_3$-$Bi_2Se_3$-$In_2Se_3$-$Bi_2Se_3$ (h).